\documentclass[twocolumn,showpacs,preprintnumbers,prc,floatfix]{revtex4}
\usepackage{amsmath,amssymb}
\usepackage{graphicx}
\usepackage{dcolumn}
\usepackage[vcentermath]{youngtab}
\usepackage[dvips]{epsfig,color}
\usepackage{bm}
\begin{document}
\title{Strange and nonstrange baryon spectra in the relativistic interacting quark-diquark model with a G\"ursey and Radicati-inspired exchange interaction}   
\author{E. Santopinto}\thanks{Corresponding author: santopinto@ge.infn.it}
\affiliation{INFN, Sezione di Genova, Via Dodecaneso 33, I-16146 Genova, Italy}
\author{J. Ferretti}
\affiliation{Dipartimento di Fisica and INFN, Universit\`a di Roma "Sapienza", Piazzale A. Moro 5, I-00185 Roma, Italy}     
\begin{abstract}
The relativistic interacting quark-diquark model, constructed in the framework of point form dynamics, is extended to strange baryons. The strange and nonstrange baryon spectra are calculated and compared with the experimental data.
\end{abstract}
\pacs{12.39.Ki, 12.39.Pn, 14.20.Gk, 14.20.Jn}
\maketitle

\section{Introduction}
The three quark constituent quark models (QMs) \cite{Isgur:1979be,Capstick:1986bm,Glozman-Riska,Loring:2001kx,Giannini:2001kb,BIL} are quite successful in describing many baryon observables, like the magnetic moments, the open-flavor decays and the electromagnetic form factors of the nucleon. 
These models show some differences, for example concerning the particular form of Hamiltonian they are based on, but share the main features: 1) they are built upon the effective degrees of freedom of three constituent quarks; 2) their mass operator contains a confining potential which, in general, is linear or quadratic in the quark relative coordinate. 
In the 80's, the predictive power of QMs has been extended with the unquenched quark model (UQM) formalism \cite{Ono:1983rd,Tornqvist,Geiger-Isgur,Bijker:2009up,Santopinto:2010zza,Bijker:2012zza,bottomonium,Ferretti:2014xqa,charmonium,Ferretti:2013vua}, which introduces the higher Fock $qqq-q \bar q$ components in baryon wave functions, arising from the coupling to the meson-baryon continuum. 
This formalism makes it possible to access to a number of problems which cannot be treated in na\"ive three quark QMs, such as the calculation of the flavor asymmetry of the nucleon sea \cite{Santopinto:2010zza} or the strange content of the nucleon electromagnetic form factors \cite{Bijker:2012zza}.

One of the main difficulties of three quarks QMs is that they predict a number of states much larger than that of the experimentally observed baryons \cite{Nakamura:2010zzi}. This is the well-known problem of the missing resonances.
One may try to look for these resonances in channels such as $N \eta$, $N \eta'$, $N \omega$ and $K^+ \Lambda$, where the final state meson is different from the pion \cite{Capstick:1992th,UNP01}. Indeed, it is well-known that the majority of baryon resonances is seen in reactions in which the pion is present either in the incoming (e.g. $N \pi \rightarrow N \pi$) or outgoing (e.g. $N \gamma \rightarrow N \pi$) channel. Thus, it would not be surprising if some baryon resonances, very weakly coupled to the single pion, were missing from the experimental results.
The other possibility is that the problem of missing resonances has to do with the choice of the effective degrees of freedom. 
Thus, in quark-diquark models, the effective degree of freedom of diquark is introduced to describe baryons as bound states of a constituent quark and diquark \cite{Ida:1966ev,lich}. 
Since the degrees of freedom of two quarks are frozen in the diquark, the state space will be greatly reduced.

The diquark concept dates back to 1964, when its possibility was hypothesized by Gell-Mann \cite{GellMann:1964nj} in his original paper on quarks.
Many articles have been written on this subject since its introduction (for a review, see Ref. \cite{Anselmino}) and more recently the diquark concept has been used in several studies, ranging from one-gluon exchange to lattice QCD calculations \cite{Jakob:1997,Brodsky:2002,Gamberg:2003,Jaffe:2003,Wilczek:2004im,Jaffe:2004ph,Selem:2006nd,DeGrand:2007vu,BacchettaRadici,Forkel:2008un,Anisovich:2010wx}. 
Important phenomenological indications for diquark-like correlations \cite{Jaffe:2004ph,Wilczek:2004im,Selem:2006nd,Close:1988br,Neubert} and for diquark confinement \cite{Bender:1996bb} have also been provided.
This makes plausibly enough to make diquarks a part of baryon wave functions. 

In this paper, we provide a mass formula for the strange and nonstrange baryon resonances within the interacting quark-diquark model, and then compute the strange and nonstrange baryon spectra.
The relativistic interacting quark-diquark model \cite{Santopinto:2004hw,Ferretti:2011zz,DeSanctis:2011zz,qD2012,Ferretti:2014kca} is constructed with the point form formalism \cite{Klink:1998zz}, which was already used to develope point form three quark QMs for baryons such as those of Refs. \cite{Pavia-Graz,Sanctis:2007zz}.
In our model \cite{Santopinto:2004hw,Ferretti:2011zz,qD2012,Ferretti:2014kca,DeSanctis:2011zz}, baryon excitations are described as two-body quark-diquark bound states.
The relative motion between the two constituents and the Hamiltonian of the model are functions of the relative coordinate $\vec r$ and its conjugate momentum $\vec q$. 
The Hamiltonian contains a direct (Coulomb + linear confining) interaction and an exchange one, depending on the spins and isospins of the quark and the diquark.
The extension of the model \cite{Santopinto:2004hw,Ferretti:2011zz,qD2012,Ferretti:2014kca,DeSanctis:2011zz} to strange and heavy (e.g. charmed and bottomed \cite{UNP02}) baryons only needs some small changes in the spin-flavor exchange potential of the model. Specifically, it requires the substitution of the previous spin and isospin-dependent exchange interaction \cite{Ferretti:2011zz} with a more general G\"ursey-Radicati \cite{Gursey:1992dc} inspired one.

%%%%%%%%%%%%%%%%%%%%%%%%%%%%%%%%%%%%%%%%%%%%%%%%%%%%%%%%%%%%%%%%%%%
\begin{table*}
\begin{tabular}{ccccccc}
\hline
\hline  \\
$M_{[n,n]}$ & $M_{\{ n,n \}}-M_{[n,n]}$ & $M_{[n,s]}-M_{[n,n]}$ & $M_{\{ n,s \}}-M_{[n,s]}$ & $M_{\{ n,s \}}-M_{\{ n,n \}}$ 
& $M_{\{ s,s \}}-M_{\{ n,s \}}$ & Source \\ \\
\hline  \\
 --   & 0.29  & --    & 0.11  & --    & --    & \cite{Wilczek:2004im} \\ 
 --   & 0.210 & --    & 0.150 & --    & --    & \cite{Jaffe:2004ph} \\
0.50  & 0.30  & --    & --    & --    & --    & \cite{Santopinto:2004hw} \\
0.60  & 0.35  & --    & --    & --    & --    & \cite{Ferretti:2011zz} \\
0.74  & 0.21  & 0.14  & 0.17  & 0.10  & 0.08  & \cite{Burden:1996nh,Hecht:2002ej} \\
0.78  & 0.28  & --    & --    & --    & --    & \cite{Roberts:2011cf} \\
0.420 & 0.520 & --    & --    & --    & --    & \cite{Schafer:1993ra} \\
0.692 & 0.330 & --    & --    & --    & --    & \cite{Cahill:1995ka} \\
0.595 & 0.205 & 0.240 & 0.140 & 0.175 & --    & \cite{Lichtenberg:1996fi} \\
0.688 & 0.202 & 0.272 & --    & --    & --    & \cite{Maris} \\
 --   & 0.360 & --    & --    & --    & --    & \cite{Orginos:2005vr} \\
 --   & 0.183 & 0.218 & 0.176 & 0.211 & --    & \cite{Chakrabarti:2010zz} \\
 --   & 0.135 & 0.201 & 0.138 & 0.204 & 0.101 & \cite{Santopinto:2011mk} \\
0.852 & 0.224 & 0.288 & 0.148 & 0.212 & 0.084 & \cite{Galata:2012xt} \\ 
0.607 & 0.356 & 0.249 & 0.360 & 0.253 & 0.136 & this work ("Fit 2") \\\\
\hline
\hline
\end{tabular}
\caption{Mass difference (in GeV) between scalar and axial-vector diquarks according to some previous studies.}
\label{tab:massediquarkaltri}
\end{table*}
%%%%%%%%%%%%%%%%%%%%%%%%%%%%%%%%%%%%%%%%

In the end, we compare our theoretical results to the experimental data \cite{Nakamura:2010zzi}. Results for the strange baryon spectrum can also be found in Refs. \cite{Capstick:1986bm,Glozman:1997ag,Bijker:2000gq,Loring:2001ky,Giannini:2005ks,Valcarce:2005rr}.

\section{Nonrelativistic quark-diquark states}
In a quark-diquark model, baryons are assumed to be composed of a constituent quark, $q$ and a constituent diquark, $Q^2$.
In the energy range we are interested into, i.e. up to 2 GeV, the diquark can be described as two correlated quarks with no internal spatial excitations, thus in $S$-wave \cite{Santopinto:2004hw,Ferretti:2011zz}. 
Then, its color-spin-flavor wave functions must be antisymmetric.
Rainbow-ladder DSE calculations confirmed that the first spatially excited diquark, the vector diquark, has a mass much larger than those of the scalar and axial-vector diquark, i.e. the ground state diquarks \cite{Burden:1996nh,Hecht:2002ej,Roberts:2011cf}.
Moreover, as we take only light baryons into account, composed of $u$, $d$, $s$ quarks, the internal group is restricted to SU$_{\mbox{sf}}$(6). 
Using the conventional notation of denoting spin by its value and flavor and color by the dimension of the representation, the quark has spin $s_2 = \frac{1}{2}$, flavor $F_2={\bf {3}}$, and color $C_2 = {\bf {3}}$. 
Since the hadron must be colorless, the diquark must transform as ${\bf {\overline{3}}}$ under SU$_{\mbox{c}}$(3) and therefore one can have only the symmetric SU$_{\mbox{sf}}$(6) representation $\mbox{{\boldmath{$21$}}}_{\mbox{sf}}$(S), containing $s_1=0$, $F_1={\bf {\overline{3}}}$, and $s_1=1$, $F_1={\bf {6}}$, i.e., the scalar and axial-vector diquarks, respectively. 

In the following, we will indicate the possible diquark states by their constituent quarks (denoted by $s$ if strange or $n$ otherwise) in square (scalar diquarks) or brace brackets (axial-vector diquarks).
The possible scalar diquark configurations are thus $[n,n]$ and $[n,s]$, while the possible axial-vector diquark configurations are $\{n,n\}$, $\{n,s\}$ and $\{s,s\}$ \cite{Jaffe:2004ph}.
For quark-diquark states, we use the notation 
\begin{subequations}
\label{eqn:notation-qD-states}
\begin{equation}
	\left| [q,q]q; \mbox{ } (\mbox{{\boldmath{$F$}}}_1,\mbox{{\boldmath{$F$}}}_2)\mbox{{\boldmath{$F$}}}; 
	\mbox{ } (t_1,t_2)T; \mbox{ } (s_1,s_2)S \right\rangle  \mbox{ }
\end{equation}
or
\begin{equation}
	\left| \{q,q\}q; \mbox{ } (\mbox{{\boldmath{$F$}}}_1,\mbox{{\boldmath{$F$}}}_2)\mbox{{\boldmath{$F$}}}; 
	\mbox{ } (t_1,t_2)T; \mbox{ } (s_1,s_2)S \right\rangle  \mbox{ },
\end{equation}
where the SU$_{\mbox{f}}(3)$ representations of the diquark, $\mbox{{\boldmath{$F$}}}_1 = {\bf {\overline{3}}}$ or ${\bf 6}$, and the quark, $\mbox{{\boldmath{$F$}}}_2 = {\bf 3}$, are coupled to the SU$_{\mbox{f}}(3)$ representation of the baryon, $\mbox{{\boldmath{$F$}}}$. Similarly, the spins (isospins) of the diquark, $s_1$ ($t_1$), and of the quark, $s_2$ ($t_2$), are coupled to the total spin (isospin) of the baryon, $S$ ($T$).
\end{subequations}
Finally, the quark-diquark basis states for $N$, $\Delta$, $\Lambda$, $\Sigma$, $\Xi$ and $\Omega$-type baryons, written in the notation of Eq. (\ref{eqn:notation-qD-states}), are given in App. \ref{qD basis}. 
See also Table \ref{tab:massediquarkaltri}, where we report some estimations of the masses of axial-vector and scalar diquarks according to some previous studies \cite{Wilczek:2004im,Jaffe:2004ph,Santopinto:2004hw,Ferretti:2011zz,Burden:1996nh,Hecht:2002ej,Roberts:2011cf,Schafer:1993ra,Cahill:1995ka,Lichtenberg:1996fi,Maris,Orginos:2005vr,Chakrabarti:2010zz,Santopinto:2011mk,Galata:2012xt}.

%%%%%%%%%%%%%%%%%%%%%%%%%%%%%%%%%%%%%%%%
\begin{table*}
\begin{tabular}{ccccccc}
\hline
\hline \\
Parameter      & Value (Fit 1)  & Value (Fit 2)  & \mbox{ } & Parameter & Value (Fit 1) & Value (Fit 2)  \\
\hline \\
$m_n$         & 200 MeV        & 159 MeV  & & $m_s$         & 550 MeV        & 213 Mev \\ 
$m_{[n,n]}$   & 600 MeV        & 607 MeV  & & $m_{[n,s]}$   & 900 MeV        & 856 MeV \\    
$m_{\{n,n\}}$ & 950 MeV        & 963 MeV  & & $m_{\{n,s\}}$ & 1200 MeV       & 1216 MeV \\
$m_{\{s,s\}}$ & 1580 MeV       & 1352 MeV & & $\tau$        & 1.20           & 1.02 \\ 
$\mu$         & 75.0 fm$^{-1}$ & 28.4 fm$^{-1}$ & & $\beta$ & 2.15 fm$^{-2}$ & 2.36 fm$^{-2}$ \\ 
$A_S$         & 350 MeV        & -436 MeV & & $A_F$         & 100 MeV        & 193 MeV \\ 
$A_I$         & 250 MeV        & 791 MeV  & & $\sigma$      & 2.30 fm$^{-1}$ & 2.25 fm$^{-1}$ \\
 $E_0$        & 141 MeV        & 150 MeV  & & $\epsilon$    & 0.37           & $-$      \\ 
$D$           & 6.13 fm$^2$    & $-$      & & $\eta$        & 11.0 fm$^{-1}$ & $-$ \\ \\
\hline
\hline
\end{tabular}
\caption{Resulting values of the model parameters. The values denoted as "Fit 1" are obtained by fitting the mass formula to nonstrange and strange baryons, those denoted as "Fit 2" are fitted to the strange sector only.}
\label{tab:ResultingParameters}
\end{table*} 
%%%%%%%%%%%%%%%%%%%%%%%%%%%%%%%%%%%%%%%%

\section{The Mass operator}
\label{The Model} 
We consider a quark-diquark system, where $\vec{r}$ and $\vec{q}$ are the relative coordinate between the two constituents and its conjugate momentum, respectively. 
The baryon rest frame mass operator we consider is 
\begin{equation}
	\begin{array}{rcl}
	M & = & E_0 + \sqrt{\vec q\hspace{0.08cm}^2 + m_1^2} + \sqrt{\vec q\hspace{0.08cm}^2 + m_2^2} 
	+ M_{\mbox{dir}}(r)  \\
	& + & M_{\mbox{ex}}(r)  
	\end{array}  \mbox{ },
	\label{eqn:H0}
\end{equation}
where $E_0$ is a constant, $M_{\mbox{dir}}(r)$ and $M_{\mbox{ex}}(r)$ respectively the direct and the exchange diquark-quark interaction, $m_1$ and $m_2$ stand for diquark and quark masses, where $m_1$ is either $m_{[q,q]}$ or $m_{\{q,q\}}$ according if the mass operator acts on a scalar or axial-vector diquark \cite{Jaffe:2004ph,Lichtenberg:1979de,deCastro:1993sr,Schafer:1993ra,Cahill:1995ka,Lichtenberg:1996fi,Burden:1996nh,Hess:1998sd,Maris,Orginos:2005vr,Wilczek:2004im,Flambaum:2005kc,Babich:2005ay,Babich:2007ah,Eichmann:2008ef,Chakrabarti:2010zz,Santopinto:2011mk,Galata:2012xt}, with $[q,q]$ = $[n,n]$ or $[n,s]$ and $\{q,q\}$ = $\{n,n\}$, $\{n,s\}$ or $\{s,s\}$.

The direct term we consider, 
\begin{equation}
  \label{eq:Vdir}
  M_{\mbox{dir}}(r)=-\frac{\tau}{r} \left(1 - e^{-\mu r}\right)+ \beta r ~~,
\end{equation}
is the sum of a Coulomb-like interaction with a cut off and a linear confinement term.
 
We also need an exchange interaction, since this is the crucial ingredient of a quark-diquark description of baryons \cite{Santopinto:2004hw,Lichtenberg:1981pp}. Thus, we consider the following G\"ursey-Radicati \cite{Gursey:1992dc} inspired interaction
\begin{equation}
	\begin{array}{rcl}
	M_{\mbox{ex}}(r) & = & \left(-1 \right)^{L + 1} \mbox{ } e^{-\sigma r} \left[ A_S \mbox{ } \vec{s}_1 
	\cdot \vec{s}_2  \right. \\ 
	& + & \left. A_F \mbox{ } \vec{\lambda}_1^f \cdot \vec{\lambda}_2^f \mbox{ } 
	+ A_I \mbox{ } \vec{t}_1 \cdot \vec{t}_2  \right]  
	\end{array}  \mbox{ },
	\label{eqn:Vexch-strange}
\end{equation}
where $\vec{s}$ and $\vec{t}$ are the spin and the isospin operators and $\vec{\lambda}^f$ the SU$_{\mbox{f}}$(3) Gell-Mann matrices. 
In the non-strange sector, we also have a contact interaction 
\begin{equation}
	\begin{array}{rcl}
	M_{\mbox{cont}} & = & \left(\frac{m_1 m_2}{E_1 E_2}\right)^{1/2+\epsilon} \frac{\eta^3 D}{\pi^{3/2}} 
	e^{-\eta^2 r^2} \mbox{ } \delta_{L,0} \delta_{s_1,1} \\ 
	& \times & \left(\frac{m_1 m_2}{E_1 E_2}\right)^{1/2+\epsilon}
	\end{array}  \mbox{ },
\end{equation}
which was introduced in the mass operator of Ref. \cite{Ferretti:2011zz} to reproduce the $\Delta-N$ mass splitting.  
It is worthwhile comparing the exchange interactions of Eq. (\ref{eqn:Vexch-strange}) and that of Ref. \cite{Ferretti:2011zz}, 
\begin{equation}
	\begin{array}{rcl}
	M_{\mbox{ex}}(r) & = & \left(-1 \right)^{L + 1} \mbox{ } e^{-\sigma r} \left[ A_S \mbox{ } \vec{s}_1 
	\cdot \vec{s}_2  \right. \\ 
	& + & \left. A_I \mbox{ } \vec{t}_1 \cdot \vec{t}_2 + 
	A_{SI} \mbox{ } (\vec{s}_1 	\cdot \vec{s}_2) (\vec{t}_1 \cdot \vec{t}_2) \right]  
	\end{array}  \mbox{ };
	\label{eqn:Vexch-old}
\end{equation}
one can notice that the spin-isospin $(\vec{s}_1 \cdot \vec{s}_2) (\vec{t}_1 \cdot \vec{t}_2)$ term of Eq. (\ref{eqn:Vexch-old}) has here been substituted with a flavor-dependent one. The isospin dependence is still necessary in Eq. (\ref{eqn:Vexch-strange}), because there are resonances which have the same quantum numbers except the isospins. These baryons, belonging to the same SU$_{\mbox{f}}$(3) representation, have different isospins that result from different combinations of the isospins of the quark and the diquark, like $\Lambda(1600)$ and $\Sigma(1193)$ (see Tables \ref{tab:Spectrum-Sigma8} and \ref{tab:Spectrum-Lambda8}). Thus, without the introduction of an isospin dependence into the exchange interaction, the previous states, $\Lambda(1600)$ and $\Sigma(1193)$, would become degenerate and lie at the same energy.

Finally, it has to be noted that in the present work all the calculations are performed without any perturbative approximation. 

The eigenfunctions of the mass operator of Eq. (\ref{eqn:H0}) can be seen as eigenstates of the mass operator with interaction in a Bakamjian-Thomas construction \cite{Pavia-Graz,BT}. The interaction is introduced by adding an interaction term to the free mass operator $M_0 = \sqrt{\vec q\hspace{0.08cm}^2 + m_1^2} + \sqrt{\vec q\hspace{0.08cm}^2 + m_2^2}$, in such a way that the interaction commutes with 
the non-interacting Lorenz generators and with the non-interacting four velocity \cite{KP}.

The dynamics is given by a point form Bakamjian-Thomas construction. Point form means that the Lorentz group is kinematic. Furthermore, since we are doing a point form Bakamjian-Thomas construction, $P = M V_0$ where $V_0$ is the noninteracting four-velocity (with eigenvalue $v$).

The general quark-diquark state, defined on the product space $H_1 \otimes H_2$ of the one-particle spin $s_1$ (0 or 1) and spin $s_2$ (1/2) positive energy representations $H_1=L^2(R^3)\otimes S_1^{0}$ or $H_1=L^2(R^3)\otimes S_1^{1}$ and $H_2=L^2(R^3)\otimes S_2^{1/2}$ of the Poincar\'e Group, is given by \cite{Ferretti:2011zz}
\begin{equation}
	\left|  p_1, p_2, \lambda_1, \lambda_2 \right\rangle \mbox{ },
\end{equation} 
where $p_1$ and $p_2$ are the four-momenta of the diquark and the quark, respectively, while $\lambda_1$ and $\lambda_2$ are, respectively, the $z$-projections of their spins.

The velocity states are introduced as \cite{Klink:1998zz,Pavia-Graz,Ferretti:2011zz} 
\begin{equation}
	\label{eqn:velocity-states}
	\vert v,\vec{k}_1,\lambda_1,\vec{k}_2,\lambda_2\rangle =
	U_{B(v)}\vert k_1,s_1,\lambda_1, k_2,s_2,\lambda_2 \rangle_{0}  \mbox{ },
\end{equation} 
where the suffix $0$ means that the diquark and the quark three-momenta $\vec {k}_1$ and $\vec{k}_2$ satisfy the condition: 
\begin{equation}
	\vec {k}_1 + \vec{k}_2=0 \mbox{ }.
\end{equation}
Following the standard rules of the point form approach, the boost operator $U_{B(v)}$ is taken as a canonical one, obtaining that the transformed four-momenta are given by $p_{1,2}=B(v)k_{1,2}$ and satisfy 
\begin{equation}
	\label{eq:pfe}
	p_1^\mu + p_2^\mu = \frac{P_N^\mu}{M_N} \left( \sqrt{\vec q\hspace{0.08cm}^2 + m_1^2} 
	+ \sqrt{\vec q\hspace{0.08cm}^2 + m_2^2} \right)  \mbox{ },
\end{equation}
where $P^\mu_N$ is the observed nucleon four-momentum and $M_N$ is its mass. 
The important point is that Eq. (\ref{eqn:velocity-states}) redefines the single particle spins. 
Since canonical boosts are applied, the conditions for a point form approach \cite{Klink:1998zz,melde} are satisfied.
Thus, the spins on the left hand state of Eq. (\ref{eqn:velocity-states}) perform the same Wigner rotations as $\vec k_1$ and $\vec k_2$, allowing to couple the spin and the orbital angular momentum as in the non relativistic case \cite{Klink:1998zz}, while the spins in the ket on the right hand of Eq. (\ref{eqn:velocity-states}) undergo the single particle Wigner rotations.

In Point form dynamics Eq. (\ref{eqn:H0}) corresponds to a good mass operator as it commutes with the Lorentz generators and with the four velocity. 
We diagonalize (\ref{eqn:H0}) in the Hilbert space spanned by the velocity states. 
Instead of the internal momenta $\vec{k_1}$ and $\vec{k_2}$, one can also use the relative momentum $\vec q$, conjugate to the relative coordinate $\vec{r} = \vec{r}_1 - \vec{r}_2$, thus considering the following velocity basis states:
\begin{equation}
	\vert v,\vec q,\lambda_1,\lambda_2 \rangle =
	U_{B(v)} \vert k_1,s_1,\lambda_1,k_2,s_2,\lambda_2 \rangle_{0} \mbox{ }.
\end{equation}

%%%%%%%%%%%%%%%%%%%%%%%%%%%%%%%%%%%%%%%%
\begin{table}[h1]
\begin{center}
\begin{tabular}{ccccccccc}
\hline
\hline \\
Resonance & Status & $M^{\mbox{exp.}}$ & $J^P$ & $L^P$ & $S$ & $s_1$ & $n_r$ & $M^{\mbox{calc.}}$ \\
 &  &       &  &  &  &  &  & (Fit 1) \\
 &  & (MeV) &  &  &  &  &  & (MeV) \\ \\
\hline \\
$N(939)$       $P_{11}$ & **** &  939         &  $\frac{1}{2}^+$ & $0^+$ &
$\frac{1}{2}$  & 0 & 0  & 939  \\
$N(1440)$      $P_{11}$ & **** &  1420 - 1470 &  $\frac{1}{2}^+$ & $0^+$ &
$\frac{1}{2}$  & 0 & 1  & 1511  \\
$N(1520)$      $D_{13}$ & **** &  1515 - 1525 &  $\frac{3}{2}^-$ & $1^-$ & 
$\frac{1}{2}$  & 0 & 0  & 1537  \\
$N(1535)$      $S_{11}$ & **** &  1525 - 1545 &  $\frac{1}{2}^-$ & $1^-$ & 
$\frac{1}{2}$  & 0 & 0  & 1537  \\
$N(1650)$      $S_{11}$ & **** &  1645 - 1670 &  $\frac{1}{2}^-$ & $1^-$ & 
$\frac{1}{2}$  & 1 & 0  & 1625  \\
$N(1675)$      $D_{15}$ & **** &  1670 - 1680 &  $\frac{5}{2}^-$ & $1^-$ & 
$\frac{3}{2}$  & 1 & 0  & 1746  \\
$N(1680)$      $F_{15}$ & **** &  1680 - 1690 &  $\frac{5}{2}^+$ & $2^+$ & 
$\frac{1}{2}$  & 0 & 0  & 1799  \\
$N(1700)$      $D_{13}$ & ***  &  1650 - 1750 &  $\frac{3}{2}^-$ & $1^-$ & 
$\frac{1}{2}$  & 1 & 0  & 1625  \\
$N(1710)$      $P_{11}$ & ***  &  1680 - 1740 &  $\frac{1}{2}^+$ & $0^+$ & 
$\frac{1}{2}$  & 1 & 0  & 1776  \\
$N(1720)$      $P_{13}$ & **** &  1700 - 1750 &  $\frac{3}{2}^+$ & $0^+$ & 
$\frac{3}{2}$  & 1 & 0  & 1648  \\
missing        &  -- & -- & $\frac{1}{2}^-$ & $1^-$ & $\frac{3}{2}$  & 1 & 0  &  1746  \\
missing        &  -- & -- & $\frac{3}{2}^-$ & $1^-$ & $\frac{3}{2}$  & 1 & 0  &  1746  \\
missing        &  -- & -- & $\frac{3}{2}^+$ & $2^+$ & $\frac{1}{2}$  & 0 & 0  &  1799  \\
$N(1875)$      $D_{13}$ & ***  &  1820 - 1920 &  $\frac{3}{2}^-$ & $1^-$ & 
$\frac{1}{2}$  & 0 & 1  & 1888  \\
$N(1880)$      $P_{11}$ & **   & 1835 - 1905  &  $\frac{1}{2}^+$ & $0^+$ &
$\frac{1}{2}$  & 0 & 2  &  1890  \\
$N(1895)$      $S_{11}$ & **   & 1880 - 1910  &  $\frac{1}{2}^-$ & $1^-$ & 
$\frac{1}{2}$  & 0 & 1  & 1888  \\
$N(1900)$ $P_{13}$      & ***  & 1875 - 1935  &  $\frac{3}{2}^+$ & $0^+$ &
$\frac{3}{2}$  & 1 & 1  &  1947  \\   \\
$\Delta(1232)$ $P_{33}$ & **** &  1230 - 1234 &  $\frac{3}{2}^+$ & $0^+$ & 
$\frac{3}{2}$  & 1 & 0  & 1247  \\
$\Delta(1600)$ $P_{33}$ & ***  &  1500 - 1700 &  $\frac{3}{2}^+$ & $0^+$ & 
$\frac{3}{2}$  & 1 & 1  & 1689  \\
$\Delta(1620)$ $S_{31}$ & **** &  1600 - 1660 &  $\frac{1}{2}^-$ & $1^-$ & 
$\frac{1}{2}$  & 1 & 0  & 1830 \\
$\Delta(1700)$ $D_{33}$ & **** &  1670 - 1750 &  $\frac{3}{2}^-$ & $1^-$ & 
$\frac{1}{2}$  & 1 & 0  & 1830 \\
$\Delta(1750)$ $P_{31}$ & *  & 1708 - 1780  &  $\frac{1}{2}^+$ & $0^+$ & 
$\frac{1}{2}$  & 1 & 0  & 1489  \\
$\Delta(1900)$ $S_{31}$ & **   &  1840 - 1920 &  $\frac{1}{2}^-$ & $1^-$ & 
$\frac{3}{2}$  & 1 & 0  & 1910 \\
$\Delta(1905)$ $F_{35}$ & **** &  1855 - 1910 &  $\frac{5}{2}^+$ & $2^+$ & 
$\frac{3}{2}$  & 1 & 0  & 2042  \\
$\Delta(1910)$ $P_{31}$ & **** &  1860 - 1920 &  $\frac{1}{2}^+$ & $2^+$ & 
$\frac{3}{2}$  & 1 & 0  & 1827  \\
$\Delta(1920)$ $P_{33}$ & ***  &  1900 - 1970 &  $\frac{3}{2}^+$ & $2^+$ & 
$\frac{3}{2}$  & 1 & 0  & 2042  \\
$\Delta(1930)$ $D_{35}$ & ***  &  1900 - 2000 &  $\frac{5}{2}^-$ & $1^-$ & 
$\frac{3}{2}$  & 1 & 0  & 1910  \\
$\Delta(1940)$ $D_{33}$ & **  &  1940 - 2060  &  $\frac{3}{2}^-$ & $1^-$ & 
$\frac{3}{2}$  & 1 & 0  & 1910 \\
$\Delta(1950)$ $F_{37}$ & **** &  1915 - 1950 &  $\frac{7}{2}^+$ & $2^+$ & 
$\frac{3}{2}$  & 1 & 0  & 2042  \\  \\
\hline
\hline
\end{tabular}
\end{center}
\caption{Comparison between the experimental \cite{Nakamura:2010zzi} values of non strange baryon resonances masses (up to 2 GeV) and the numerical ones, from "Fit 1". $J^P$ and $L^P$ are respectively the total angular momentum and the orbital angular momentum of the baryon, including the parity $P$; $S$ is the total spin, obtained coupling the spin of the diquark, $s_1$, and that of the quark; finally $n_r$ is the number of nodes in the radial wave function. Since in the nonstrange sector we can only have two type of diquarks, the scalar, $[n,n]$, and axial-vector diquark, $\{n,n\}$, with spin $s_1 = 0$ and 1, respectively, for simplicity here we use the notation of Refs. \cite{Ferretti:2011zz,qD2012}.} 
\label{tab:Spectrum-NDelta-Fit1}
\end{table}
%%%%%%%%%%%%%%%%%%%%%%%%%%%%%%%%%%%%%%%%

%%%%%%%%%%%%%%%%%%%%%%%%%%%%%%%%%%%%%%%%
\begin{table*}
\begin{tabular}{cccccccccccccc}
\hline
\hline \\
Resonance & Status & $M^{\mbox{exp.}}$ & $J^P$ & $L^P$ & $S$ & $s_1$ & $Q^2q$ & ${\bf F}$ & ${\bf {F_1}}$ & $I$ & $t_1$ & $n_r$ & $M^{\mbox{calc.}}$ (Fit 1) \\
 &  & (MeV) &  &  &  &  &  &  & & & & & (MeV) \\ \\
\hline \\
$\Sigma(1193)$ $P_{11}$   & **** & 1189 - 1197    & $\frac{1}{2}^+$ & $0^+$ & $\frac{1}{2}$ & 0 & $[n,s]n$   & ${\bf 8}$ & ${\bf {\bar 3}}$ & 1 & $\frac{1}{2}$ & 0 & 1134 \\
$\Sigma(1660)$ $P_{11}$   & ***  & 1630 - 1690    & $\frac{1}{2}^+$ & $0^+$ & $\frac{1}{2}$ & 1 & $\{n,s\}n$ & ${\bf 8}$ & ${\bf 6}$        & 1 & $\frac{1}{2}$ & 0 & 1734 \\
$\Sigma(1670)$ $D_{13}$   & **** & 1665 - 1685    & $\frac{3}{2}^-$ & $1^-$ & $\frac{1}{2}$ & 0 & $[n,s]n$ & ${\bf 8}$ & ${\bf {\bar 3}}$ & 1 & $\frac{1}{2}$ & 0 & 1800 \\
$\Sigma(1750)$ $S_{11}$   & ***  & 1730 - 1800    & $\frac{1}{2}^-$ & $1^-$ & $\frac{1}{2}$ & 0 & $[n,s]n$   & ${\bf 8}$ & ${\bf {\bar 3}}$ & 1 & $\frac{1}{2}$ & 0 & 1800 \\
$\Sigma(1770)$ $P_{11}$   & *    & $\approx1770$  & $\frac{1}{2}^+$ & $0^+$ & $\frac{1}{2}$ & 0 & $[n,s]n$ & ${\bf 8}$ & ${\bf {\bar 3}}$ & 1 & $\frac{1}{2}$ & 1 & 1739 \\
$\Sigma(1775)$ $D_{15}$   & **** & 1770 - 1780    & $\frac{5}{2}^-$ & $1^-$ & $\frac{3}{2}$ & 1 & $\{n,s\}n$ & ${\bf 8}$ & ${\bf 6}$     & 1 & $\frac{1}{2}$ & 0 & 2030 \\
 missing                  & --   & --             & $\frac{1}{2}^-$ & $1^-$ & $\frac{1}{2}$ & 1 & $\{n,n\}s$ & ${\bf 8}$ & ${\bf 6}$        & 1 & 1             & 0 & 1872 \\
 missing                  & --   & --             & $\frac{3}{2}^-$ & $1^-$ & $\frac{1}{2}$ & 1 & $\{n,n\}s$ & ${\bf 8}$ & ${\bf 6}$        & 1 & 1             & 0 & 1872 \\
$\Sigma(1880)$ $P_{11}$   & **   & $\approx1880$  & $\frac{1}{2}^+$ & $0^+$ & $\frac{1}{2}$ & 1 & $\{n,n\}s$   & ${\bf 8}$ & ${\bf 6}$        & 1 & 1           & 0 & 1751 \\
$\Sigma(1915)$ $F_{15}$   & **** & 1900 - 1935    & $\frac{5}{2}^+$ & $2^+$ & $\frac{1}{2}$ & 0 & $[n,s]n$   & ${\bf 8}$ & ${\bf {\bar 3}}$ & 1 & $\frac{1}{2}$ & 0 & 2041 \\
$\Sigma(1940)$ $D_{13}$   & ***  & 1900 - 1950    & $\frac{3}{2}^-$ & $1^-$ & $\frac{1}{2}$ & 1 & $\{n,s\}n$   & ${\bf 8}$ & ${\bf 6}$        & 1 & $\frac{1}{2}$ & 0 & 1916 \\ 
$\Sigma(2000)$ $S_{11}$   & *    & $\approx2000$  & $\frac{1}{2}^-$ & $1^-$ & $\frac{1}{2}$ & 1 & $\{n,s\}n$ & ${\bf 8}$ & ${\bf 6}$        & 1 & $\frac{1}{2}$ & 0 & 1916 \\   \\ 
$\Xi(1318)$ $P_{11}$     & **** & 1315 - 1322    & $\frac{1}{2}^+$ & $0^+$ & $\frac{1}{2}$ & 0 & $[n,s]s$   & ${\bf 8}$ & ${\bf {\bar 3}}$ & $\frac{1}{2}$ & $\frac{1}{2}$ & 0 & 1343  \\
$\Xi(1820)$ $D_{13}$     & ***  & 1818 - 1828    & $\frac{3}{2}^-$ & $1^-$ & $\frac{1}{2}$ & 0 & $[n,s]s$   & ${\bf 8}$ & ${\bf {\bar 3}}$ & $\frac{1}{2}$ & $\frac{1}{2}$ & 0 & 2002  \\ 
 missing                  & --   & --            & $\frac{1}{2}^+$ & $0^+$ & $\frac{1}{2}$ & 1 & $\{n,s\}s$ & ${\bf 8}$ & ${\bf 6}$        & $\frac{1}{2}$ & $\frac{1}{2}$ & 0 & 1965  \\
 missing                  & --   & --            & $\frac{1}{2}^+$ & $0^+$ & $\frac{1}{2}$ & 0 & $[n,s]s$   & ${\bf 8}$ & ${\bf {\bar 3}}$ & $\frac{1}{2}$ & $\frac{1}{2}$ & 1 & 1978  \\  \\
$\Lambda(1116)$ $P_{01}$ & **** & 1116        & $\frac{1}{2}^+$ & $0^+$ & $\frac{1}{2}$ & 0   & $[n,n]s$   & ${\bf 8}$ & ${\bf {\bar 3}}$ & 0 & 0             & 0 & 1128   \\
$\Lambda(1600)$ $P_{01}$ & ***  & 1560 - 1700 & $\frac{1}{2}^+$ & $0^+$ & $\frac{1}{2}$ & 0   & $[n,s]n$   & ${\bf 8}$ & ${\bf {\bar 3}}$ & 0 & $\frac{1}{2}$ & 0 & 1256   \\
 missing                 & --   & --          & $\frac{3}{2}^+$ & $0^+$ & $\frac{3}{2}$ & 1   & $\{n,s\}n$ & ${\bf 8}$ & ${\bf {6}}$      & 0 & $\frac{1}{2}$ & 0 & 1613   \\
$\Lambda(1670)$ $S_{01}$ & **** & 1660 - 1680 & $\frac{1}{2}^-$ & $1^-$ & $\frac{1}{2}$ & 0   & $[n,n]s$   & ${\bf 8}$ & ${\bf {\bar 3}}$ & 0 & 0             & 0 & 1756   \\
$\Lambda(1690)$ $D_{03}$ & **** & 1685 - 1695 & $\frac{3}{2}^-$ & $1^-$ & $\frac{1}{2}$ & 0   & $[n,n]s$   & ${\bf 8}$ & ${\bf {\bar 3}}$ & 0 & 0             & 0 & 1756  \\
 missing                 & --   & --          & $\frac{1}{2}^+$ & $0^+$ & $\frac{1}{2}$ & 0   & $[n,n]s$   & ${\bf 8}$ & ${\bf {\bar 3}}$ & 0 & 0             & 1 & 1738   \\
 missing                 & --   & --          & $\frac{1}{2}^-$ & $1^-$ & $\frac{1}{2}$ & 0   & $[n,s]n$   & ${\bf 8}$ & ${\bf {\bar 3}}$ & 0 & $\frac{1}{2}$ & 0 & 1758   \\
 missing                 & --   & --          & $\frac{3}{2}^-$ & $1^-$ & $\frac{1}{2}$ & 0   & $[n,s]n$   & ${\bf 8}$ & ${\bf {\bar 3}}$ & 0 & $\frac{1}{2}$ & 0 & 1758   \\	
$\Lambda(1800)$ $S_{01}$ & ***  & 1720 - 1850 & $\frac{1}{2}^-$ & $1^-$ & $\frac{3}{2}$ & 1   & $\{n,s\}n$ & ${\bf 8}$ & ${\bf {6}}$      & 0 & $\frac{1}{2}$ & 0 & 1853   \\
$\Lambda(1810)$ $P_{01}$ & ***  & 1750 - 1850 & $\frac{1}{2}^+$ & $0^+$ & $\frac{1}{2}$ & 0   & $[n,s]n$   & ${\bf 8}$ & ${\bf {\bar 3}}$ & 0 & $\frac{1}{2}$ & 1 & 1794   \\
$\Lambda(1820)$ $F_{05}$ & **** & 1815 - 1825 & $\frac{5}{2}^+$ & $2^+$ & $\frac{1}{2}$ & 0   & $[n,n]s$   & ${\bf 8}$ & ${\bf {\bar 3}}$ & 0 & 0             & 0 & 2006   \\
$\Lambda(1830)$ $D_{05}$ & **** & 1810 - 1830 & $\frac{5}{2}^-$ & $1^-$ & $\frac{3}{2}$ & 1   & $\{n,s\}n$ & ${\bf 8}$ & ${\bf {6}}$      & 0 & $\frac{1}{2}$ & 0 & 1979   \\
 missing                 & --   & --          & $\frac{1}{2}^+$ & $0^+$ & $\frac{1}{2}$ & 1   & $\{n,s\}n$ & ${\bf 8}$ & ${\bf {6}}$      & 0 & $\frac{1}{2}$ & 0 & 1832   \\
  missing                 & --   & --          & $\frac{3}{2}^-$ & $1^-$ & $\frac{3}{2}$ & 1   & $\{n,s\}n$ & ${\bf 8}$ & ${\bf {6}}$      & 0 & $\frac{1}{2}$ & 0 & 1853   \\
$\Lambda(1890)$ $P_{03}$ & **** & 1850 - 1910 & $\frac{3}{2}^+$ & $2^+$ & $\frac{1}{2}$ & 0   & $[n,n]s$ & ${\bf 8}$ & ${\bf {\bar 3}}$ & 0 & 0 & 0 & 2006   \\ 
 missing                 & --   & --          & $\frac{1}{2}^-$ & $1^-$ & $\frac{3}{2}$ & 1   & $\{n,s\}n$ & ${\bf 8}$ & ${\bf {6}}$      & 0 & $\frac{1}{2}$ & 0 & 1979   \\
 missing                 & --   & --          & $\frac{3}{2}^-$ & $1^-$ & $\frac{3}{2}$ & 1   & $\{n,s\}n$ & ${\bf 8}$ & ${\bf {6}}$      & 0 & $\frac{1}{2}$ & 0 & 1979   \\  \\
\hline
\hline
\end{tabular}
\caption{Comparison between the experimental values \cite{Nakamura:2010zzi} of $\Sigma$, $\Xi$ and $\Lambda$-type resonance masses (up to 2 GeV) and the numerical ones (all values are expressed in $MeV$), from "Fit 1". $J^P$ and $L^P$ are respectively the total angular momentum and the orbital angular momentum of the baryon, including the parity $P$; $S$ is the total spin, obtained by coupling the spin of the diquark $s_1$ and that of the quark; $Q^2q$ stands for the diquark-quark structure of the state; ${\bf F}$ and ${\bf {F_1}}$ are the dimensions of the SU$_{\mbox{f}}$(3) representations for the baryon and the diquark, respectively; $I$ and $t_1$ are the isospins of the baryon and the diquark, respectively; finally $n_r$ is the number of nodes in the radial wave function.} 
\label{tab:Spectrum-Sigma8-Fit1}
\end{table*}  
%%%%%%%%%%%%%%%%%%%%%%%%%%%%%%%%%%%%%%%%

\section{Results and discussion}
\begin{figure}[htbp]
\begin{center}
\includegraphics[width=7cm]{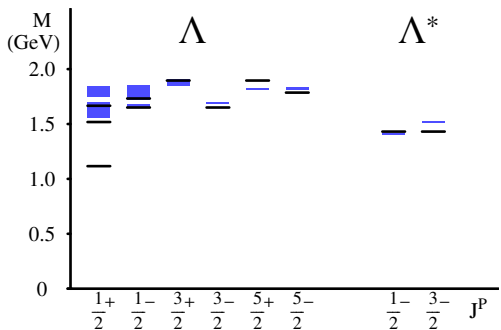}
\end{center}
\caption{(Color online) Comparison between the calculated masses (black lines) of the $3^*$ and $4^*$ $\Lambda$ and $\Lambda^*$ resonances (up to 2 GeV; from "Fit 2") and the experimental masses from PDG \cite{Nakamura:2010zzi} (blue boxes).} 
\label{fig:Spectrum-lambda}
\end{figure}

\begin{figure}[htbp]
\begin{center}
\includegraphics[width=7cm]{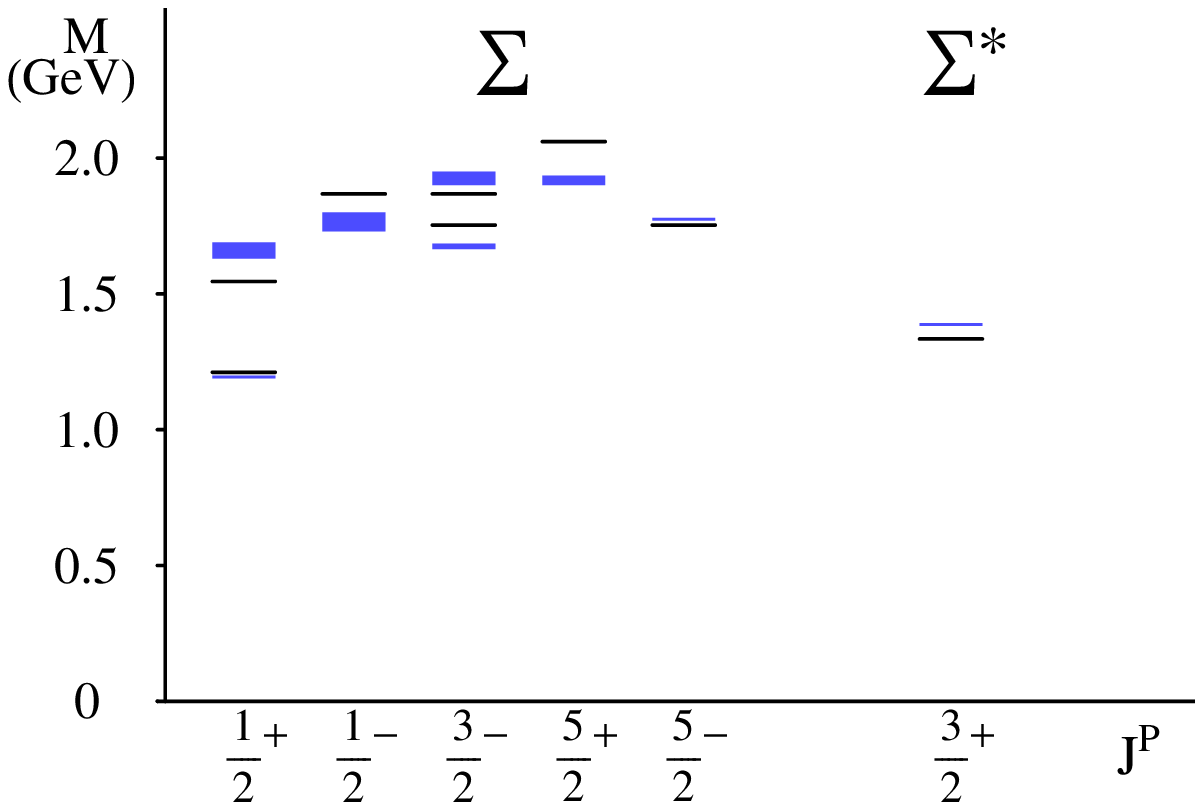}
\end{center}
\caption{(Color online) Comparison between the calculated masses (black lines) of the $3^*$ and $4^*$ $\Sigma$ and $\Sigma^*$ resonances (up to 2 GeV; from "Fit 2") and the experimental masses from PDG \cite{Nakamura:2010zzi} (blue boxes).} 
\label{fig:Spectrum-sigma}
\end{figure}

\begin{figure}[htbp]
\begin{center}
\includegraphics[width=7cm]{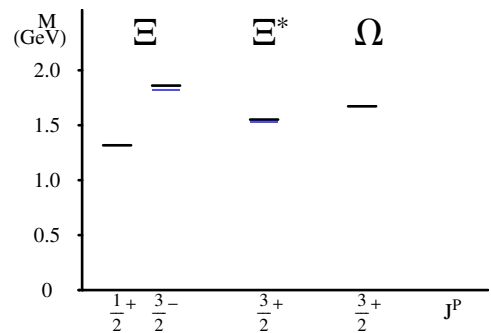}
\end{center}
\caption{(Color online) Comparison between the calculated masses (black lines) of the $3^*$ and $4^*$ $\Xi$, $\Xi^*$ and $\Omega$ resonances (up to 2 GeV; from "Fit 2") and the experimental masses from PDG \cite{Nakamura:2010zzi} (blue boxes).} 
\label{fig:Spectrum-xi-omega}
\end{figure}

In this section, we show our results for the strange and non-strange baryon spectra.
Because this paper is mainly focused on the extension of the interacting quark-diquark model to strange baryons, here we present the results of two fits to the experimental data \cite{Nakamura:2010zzi}. In the first, "Fit 1", we fit the model mass formula to the strange and non-strange baryon spectra, while in the second, "Fit 2", we focus our attention on the strange sector only.  
Obviously, in this second case we expect to get a better reproduction of the experimental data in the strange baryon sector and, perhaps, to increase the predictive power of our model for still unobserved strange baryon resonances. 
Tables \ref{tab:Spectrum-NDelta-Fit1} and \ref{tab:Spectrum-Sigma8-Fit1} show the comparison between the experimental data and the results of our quark-diquark model calculation, obtained with the set of parameters of Table \ref{tab:ResultingParameters} ("Fit 1") and a rms of 119 MeV.
Figures \ref{fig:Spectrum-lambda}-\ref{fig:Spectrum-xi-omega} and Tables \ref{tab:Spectrum-Sigma8}-\ref{tab:Spectrum-Lambda8} show our quark-diquark model results, obtained with the set of parameters of Table \ref{tab:ResultingParameters} ("Fit 2") and a rms of 72 MeV.
Our results can be compared to those of Refs. \cite{Glozman:1997ag,Bijker:2000gq,Loring:2001ky,Capstick:1986bm,Giannini:2005ks,Valcarce:2005rr}.  

There is a certain difference between the values of the model parameters used in the two fits. This is especially evident in the case of the quark masses and the exchange potential parameters. The values of the parameters strongly depend from one another. Thus, e.g. if we modify those for the exchange potential, this will also have an effect on the constituent quark masses. Moreover, and most important, some parameters are present in the first fit and not in second, because they were introduced in the non strange sector to reproduce the $\Delta-N$ mass splitting, and thus they are inessential in the strange sector. In fact, we can say that the non-strange sector is a special case. This is because spin forces are stronger in this sector than in the others. This can be seen not only in baryons, but also in meson spectroscopy, where the pion mass results from very large hyperfine contributions, while, for example, in the strange or charmed sectors spin forces are much weaker. This is the reason why we expect to get better results for heavy baryons \cite{FS-inprep}, where spin-forces are weaker and can be treated more easily.

A long standing problem of three quarks QMs in the strange sector is that of $\Lambda^*(1405)$, since its experimental mass is not reproduced with a reasonable accuracy within this kind of models.
Here, the mass of this resonance is well reproduced in terms of a quark-diquark picture of baryons. 
It is also interesting to note that in our model $\Lambda(1116)$ and $\Lambda^*(1520)$ are described as bound states of a scalar diquark $[n,n]$ and a quark $s$, where the quark-diquark system is in $S$ or $P$-wave, respectively. This is in accordance with the observations of Refs. \cite{Jaffe:2004ph,Selem:2006nd} on $\Lambda$'s fragmentation functions, that the two resonances can be described as $[n,n]-s$ systems. See Table \ref{tab:Spectrum-Lambda8}.

The presence of more diquark types, with respect to the non-strange case of Ref. \cite{Ferretti:2011zz}, makes the reproduction of the experimental data below the energy of 2 GeV more difficult than before.  
In particular, one can notice that in the present case (see results from "Fit 2", Tables \ref{tab:Spectrum-Sigma8}-\ref{tab:Spectrum-Lambda8}) there are 19 missing resonances below the energy of 2 GeV, while in the non strange sector \cite{Ferretti:2011zz} there were no missing states under 2 GeV. 
Indeed, in the strange sector one has two scalar diquarks, $[n,n]$ and $[n,s]$, and three axial-vector diquarks, $\{n,n\}$, $\{n,s\}$ and $\{s,s\}$, while in the non-strange sector one only has a scalar diquark, $[n,n]$, and an axial-vector diquark, $\{n,n\}$.
Nevertheless, we think that the number of missing resonances of our model may decrease when new experimental data from more powerful experiments and more precise data analyses are extracted. 
The search for these resonances should be one of the main goals of the baryon research programs at JLab, BES, ELSA, Crystal Barrel and TAPS. 
See also the latest multi-channel Bonn-Gatchina partial wave analysis results, including data from Crystal Barrel/TAPS at ELSA and other labs \cite{Anisovich:2011fc}.

Baryon resonances problems have already been treated with other quark-diquark models \cite{Galata:2012xt}, unquenched quark models \cite{Ono:1983rd,Tornqvist,Geiger-Isgur,Bijker:2009up,Santopinto:2010zza,Bijker:2012zza,bottomonium,Ferretti:2014xqa,charmonium,Ferretti:2013vua}, and hypercentral models \cite{Ferraris:1995ui}, but in the end baryon resonances still remain an open problem \cite{Capstick:2007tv}. 
In three quarks QMs for baryons, light baryons are ordered according to the approximate SU$_{\mbox{f}}$(3) symmetry. 
Nevertheless, on one hand many unseen excited resonances are predicted by every three-quark models; on the other hand, states with certain quantum numbers appear in the spectrum at excitation energies much lower than predicted \cite{Nakamura:2010zzi}. 
For example, in the non-strange sector up to an excitation energy of 2.41 GeV, on average about 45 $N$ states are predicted, but only 12 have been established (four- or three-star) and 7 are tentative (two- or one-star) \cite{Nakamura:2010zzi}.
A possible solution to the puzzle of missing resonances is the introduction of a new effective degree of freedom: the diquark. This is what we tried to do in the present paper and in Ref. \cite{Ferretti:2011zz} in the non-strange sector.

%%%%%%%%%%%%%%%%%%%%%%%%%%%%%%%%%%%%%%%%
%%%%%%%%%%%%%%%%%%%%%%%%%%%%%%%%%%%%%%%%
%%%%%%%%%%%%%%%%%%%%%%%%%%%%%%%%%%%%%%%%%%%%%%%%%%
%%%%%%%%%%%%%%%%%%%%%%%%%%%%%%%%%%%%%%%%%%%%%%%%%%

While the absolute values of the diquark masses are model dependent, their difference is not. Comparing our result for the mass difference between the axial-vector and scalar diquarks to those of Tab. \ref{tab:massediquarkaltri}, it is interesting to note that our estimations are comparable with the other ones. The main deviation from the evaluations reported in the table arises in the difference $\{ n,s \}-[ n,s ]$.

The whole mass operator of Eq. (\ref{eqn:H0}) has been diagonalized by means of a numerical variational procedure, based on harmonic oscillator trial wave functions. With a variational basis of 100 harmonic oscillator shells, the results converge very well.

The present work can be expanded to include charmed and/or bottomed baryons \cite{FS-inprep}, which can be quite interesting in light of the recent experimental effort to study the properties of heavy hadrons. The application of our model to the description of heavy baryons is straightforward and does not require a modification of the mass operator. 

%%%%%%%%%%%%%%%%%%%%%%%%%%%%%%%%%%%%%%%%
\begin{table*}
\begin{tabular}{cccccccccccccc}
\hline
\hline \\
Resonance & Status & $M^{\mbox{exp.}}$ & $J^P$ & $L^P$ & $S$ & $s_1$ & $Q^2q$ & ${\bf F}$ & ${\bf {F_1}}$ & $I$ & $t_1$ & $n_r$ & $M^{\mbox{calc.}}$ (Fit 2) \\
 &  & (MeV) &  &  &  &  &  &  & & & & & (MeV) \\ \\
\hline \\
$\Sigma(1193)$ $P_{11}$   & **** & 1189 - 1197    & $\frac{1}{2}^+$ & $0^+$ & $\frac{1}{2}$ & 0 & $[n,s]n$   & ${\bf 8}$ & ${\bf {\bar 3}}$ & 1 & $\frac{1}{2}$ & 0 & 1211  \\
$\Sigma(1620)$ $S_{11}$   & **   & $\approx1620$  & $\frac{1}{2}^-$ & $1^-$ & $\frac{3}{2}$ & 1 & $\{n,n\}s$ & ${\bf 8}$ & ${\bf 6}$        & 1 & 1             & 0 & 1753  \\
$\Sigma(1660)$ $P_{11}$   & ***  & 1630 - 1690    & $\frac{1}{2}^+$ & $0^+$ & $\frac{1}{2}$ & 1 & $\{n,n\}s$ & ${\bf 8}$ & ${\bf 6}$        & 1 & 1             & 0 & 1546  \\
$\Sigma(1670)$ $D_{13}$   & **** & 1665 - 1685    & $\frac{3}{2}^-$ & $1^-$ & $\frac{3}{2}$ & 1 & $\{n,n\}s$ & ${\bf 8}$ & ${\bf 6}$        & 1 & 1             & 0 & 1753  \\
$\Sigma(1750)$ $S_{11}$   & ***  & 1730 - 1800    & $\frac{1}{2}^-$ & $1^-$ & $\frac{1}{2}$ & 0 & $[n,s]n$   & ${\bf 8}$ & ${\bf {\bar 3}}$ & 1 & $\frac{1}{2}$ & 0 & 1868  \\
$\Sigma(1770)$ $P_{11}$   & *    & $\approx1770$  & $\frac{1}{2}^+$ & $0^+$ & $\frac{1}{2}$ & 1 & $\{n,s\}n$ & ${\bf 8}$ & ${\bf 6}$        & 1 & $\frac{1}{2}$ & 0 & 1668  \\
$\Sigma(1775)$ $D_{15}$   & **** & 1770 - 1780    & $\frac{5}{2}^-$ & $1^-$ & $\frac{3}{2}$ & 1 & $\{n,n\}s$ & ${\bf 8}$ & ${\bf 6}$        & 1 & 1             & 0 & 1753  \\
$\Sigma(1880)$ $P_{11}$   & **   & $\approx1880$  & $\frac{1}{2}^+$ & $0^+$ & $\frac{1}{2}$ & 0 & $[n,s]n$   & ${\bf 8}$ & ${\bf {\bar 3}}$ & 1 & $\frac{1}{2}$ & 1 & 1801  \\
$\Sigma(1915)$ $F_{15}$   & **** & 1900 - 1935    & $\frac{5}{2}^+$ & $2^+$ & $\frac{1}{2}$ & 0 & $[n,s]n$   & ${\bf 8}$ & ${\bf {\bar 3}}$ & 1 & $\frac{1}{2}$ & 0 & 2061  \\
$\Sigma(1940)$ $D_{13}$   & ***  & 1900 - 1950    & $\frac{3}{2}^-$ & $1^-$ & $\frac{1}{2}$ & 0 & $[n,s]n$   & ${\bf 8}$ & ${\bf {\bar 3}}$ & 1 & $\frac{1}{2}$ & 0 & 1868  \\ 
 missing                  & --   & --             & $\frac{3}{2}^-$ & $1^-$ & $\frac{3}{2}$ & 1 & $\{n,n\}s$ & ${\bf 8}$ & ${\bf 6}$        & 1 & 1             & 0 & 1895  \\
$\Sigma(2000)$ $S_{11}$   & *    & $\approx2000$  & $\frac{1}{2}^-$ & $1^-$ & $\frac{3}{2}$ & 1 & $\{n,n\}s$ & ${\bf 8}$ & ${\bf 6}$        & 1 & 1             & 0 & 1895  \\   \\ 
$\Sigma^*(1385)$ $P_{13}$ & **** & 1382 - 1388    & $\frac{3}{2}^+$ & $0^+$ & $\frac{3}{2}$ & 1 & $\{n,n\}s$ & ${\bf 10}$ & ${\bf 6}$       & 1 & 1             & 0 & 1334  \\
$\Sigma^*(1840)$ $P_{13}$ & *    & $\approx1840$  & $\frac{3}{2}^+$ & $0^+$ & $\frac{3}{2}$ & 1   & $\{n,s\}n$ & ${\bf 10}$ & ${\bf 6}$       & 1 & $\frac{1}{2}$ & 0 & 1439  \\
$\Sigma^*(2080)$ $P_{13}$ & **   & $\approx2080$ & $\frac{3}{2}^+$ & $0^+$ & $\frac{3}{2}$  & 1   & $\{n,n\}s$ & ${\bf 10}$ & ${\bf 6}$       & 1 & 1             & 1 & 1924  \\  \\
\hline
\hline
\end{tabular}
\caption{Comparison between the experimental values \cite{Nakamura:2010zzi} of $\Sigma$ and $\Sigma^*$-type resonance masses (up to 2 GeV) and the numerical ones (all values are expressed in $MeV$), from "Fit 2". $J^P$ and $L^P$ are respectively the total angular momentum and the orbital angular momentum of the baryon, including the parity $P$; $S$ is the total spin, obtained by coupling the spin of the diquark $s_1$ and that of the quark; $Q^2q$ stands for the diquark-quark structure of the state; ${\bf F}$ and ${\bf {F_1}}$ are the dimensions of the SU$_{\mbox{f}}$(3) representations for the baryon and the diquark, respectively; $I$ and $t_1$ are the isospins of the baryon and the diquark, respectively; finally $n_r$ is the number of nodes in the radial wave function.} 
\label{tab:Spectrum-Sigma8}
\end{table*}  
%%%%%%%%%%%%%%%%%%%%%%%%%%%%%%%%%%%%%%%%

%%%%%%%%%%%%%%%%%%%%%%%%%%%%%%%%%%%%%%%%
\begin{table*}
\begin{tabular}{cccccccccccccc}
\hline
\hline \\
Resonance & Status & $M^{\mbox{exp.}}$ & $J^P$ & $L^P$ & $S$ & $s_1$ & $Q^2q$ & ${\bf F}$ & ${\bf {F_1}}$ & $I$ & $t_1$ & $n_r$ & $M^{\mbox{calc.}}$ (Fit 2) \\
 &  & (MeV) &  &  &  &  &  &  & & & & & (MeV) \\ \\
\hline \\
$\Xi(1318)$ $P_{11}$     & **** & 1315 - 1322    & $\frac{1}{2}^+$ & $0^+$ & $\frac{1}{2}$ & 0 & $[n,s]s$   & ${\bf 8}$ & ${\bf {\bar 3}}$ & $\frac{1}{2}$ & $\frac{1}{2}$ & 0 & 1317  \\
 missing                  & --   & --            & $\frac{1}{2}^+$ & $0^+$ & $\frac{1}{2}$ & 1 & $\{n,s\}s$ & ${\bf 8}$ & ${\bf 6}$        & $\frac{1}{2}$ & $\frac{1}{2}$ & 0 & 1772  \\
$\Xi(1820)$ $D_{13}$     & ***  & 1818 - 1828    & $\frac{3}{2}^-$ & $1^-$ & $\frac{1}{2}$ & 0 & $[n,s]s$   & ${\bf 8}$ & ${\bf {\bar 3}}$ & $\frac{1}{2}$ & $\frac{1}{2}$ & 0 & 1861  \\ 
 missing                  & --   & --            & $\frac{1}{2}^+$ & $0^+$ & $\frac{1}{2}$ & 0 & $[n,s]s$   & ${\bf 8}$ & ${\bf {\bar 3}}$ & $\frac{1}{2}$ & $\frac{1}{2}$ & 1 & 1868  \\
 missing                  & --   & --            & $\frac{1}{2}^+$ & $0^+$ & $\frac{1}{2}$ & 1 & $\{s,s\}n$ & ${\bf 8}$ & ${\bf 6}$        & $\frac{1}{2}$ & 0         & 0 & 1874  \\
 missing                  & --   & --            & $\frac{3}{2}^-$ & $1^-$ & $\frac{3}{2}$ & 1 & $\{n,s\}s$ & ${\bf 8}$ & ${\bf 6}$        & $\frac{1}{2}$ & $\frac{1}{2}$ & 0 & 1971  \\  \\
$\Xi^*(1530)$ $P_{13}$   & **** & 1531 - 1532    & $\frac{3}{2}^+$ & $0^+$ & $\frac{3}{2}$  & 1 & $\{n,s\}s$ & ${\bf 10}$ & ${\bf 6}$       & $\frac{1}{2}$ & $\frac{1}{2}$ & 0 & 1552  \\
 missing                  & --   & --            & $\frac{3}{2}^+$ & $0^+$ & $\frac{3}{2}$  & 1 & $\{s,s\}n$ & ${\bf 10}$ & ${\bf 6}$       & $\frac{1}{2}$ & 0             & 0 & 1653  \\ \\
 $\Omega(1672)$ $P_{03}$     & **** & 1672 - 1673    & $\frac{3}{2}^+$ & $0^+$ & $\frac{3}{2}$ & 1 & $\{s,s\}s$ & ${\bf 10}$ & ${\bf 6}$ & 0 & 0 & 0 & 1672  \\ \\
\hline
\hline
\end{tabular}
\caption{As table \ref{tab:Spectrum-Sigma8}, but for $\Xi$, $\Xi^*$ and $\Omega$-type resonances.} 
\label{tab:Spectrum-Xi8} 
\end{table*}
%%%%%%%%%%%%%%%%%%%%%%%%%%%%%%%%%%%%%%%%

%%%%%%%%%%%%%%%%%%%%%%%%%%%%%%%%%%%%%%%%
\begin{table*}
\begin{tabular}{cccccccccccccc}
\hline
\hline \\
Resonance & Status & $M^{\mbox{exp.}}$ & $J^P$ & $L^P$ & $S$ & $s_1$ & $Q^2q$ & ${\bf F}$ & ${\bf {F_1}}$ & $I$ & $t_1$ & $n_r$ & $M^{\mbox{calc.}}$ (Fit 2) \\
 &  & (MeV) &  &  &  &  &  &  & & & & & (MeV) \\ \\
\hline \\
$\Lambda(1116)$ $P_{01}$ & **** & 1116        & $\frac{1}{2}^+$ & $0^+$ & $\frac{1}{2}$ & 0   & $[n,n]s$   & ${\bf 8}$ & ${\bf {\bar 3}}$ & 0 & 0             & 0 & 1116  \\
$\Lambda(1600)$ $P_{01}$ & ***  & 1560 - 1700 & $\frac{1}{2}^+$ & $0^+$ & $\frac{1}{2}$ & 0   & $[n,s]n$   & ${\bf 8}$ & ${\bf {\bar 3}}$ & 0 & $\frac{1}{2}$ & 0 & 1518  \\
$\Lambda(1670)$ $S_{01}$ & **** & 1660 - 1680 & $\frac{1}{2}^-$ & $1^-$ & $\frac{1}{2}$ & 0   & $[n,n]s$   & ${\bf 8}$ & ${\bf {\bar 3}}$ & 0 & 0             & 0 & 1650  \\
$\Lambda(1690)$ $D_{03}$ & **** & 1685 - 1695 & $\frac{3}{2}^-$ & $1^-$ & $\frac{1}{2}$ & 0   & $[n,n]s$   & ${\bf 8}$ & ${\bf {\bar 3}}$ & 0 & 0             & 0 & 1650  \\
 missing                 & --   & --          & $\frac{3}{2}^-$ & $1^-$ & $\frac{1}{2}$ & 0   & $[n,s]n$   & ${\bf 8}$ & ${\bf {\bar 3}}$ & 0 & $\frac{1}{2}$ & 0 & 1732  \\
 missing                 & --   & --          & $\frac{1}{2}^-$ & $1^-$ & $\frac{3}{2}$ & 1   & $\{n,s\}n$ & ${\bf 8}$ & ${\bf {6}}$      & 0 & $\frac{1}{2}$ & 0 & 1785  \\
  missing                 & --   & --          & $\frac{3}{2}^-$ & $1^-$ & $\frac{1}{2}$ & 0   & $[n,n]s$   & ${\bf 8}$ & ${\bf {\bar 3}}$ & 0 & 0             & 1 & 1785  \\
$\Lambda(1800)$ $S_{01}$ & ***  & 1720 - 1850 & $\frac{1}{2}^-$ & $1^-$ & $\frac{1}{2}$ & 0   & $[n,s]n$   & ${\bf 8}$ & ${\bf {\bar 3}}$ & 0 & $\frac{1}{2}$ & 0 & 1732  \\
$\Lambda(1810)$ $P_{01}$ & ***  & 1750 - 1850 & $\frac{1}{2}^+$ & $0^+$ & $\frac{1}{2}$ & 0   & $[n,n]s$   & ${\bf 8}$ & ${\bf {\bar 3}}$ & 0 & 0             & 1 & 1666  \\
$\Lambda(1820)$ $F_{05}$ & **** & 1815 - 1825 & $\frac{5}{2}^+$ & $2^+$ & $\frac{1}{2}$ & 0   & $[n,n]s$   & ${\bf 8}$ & ${\bf {\bar 3}}$ & 0 & 0             & 0 & 1896  \\
$\Lambda(1830)$ $D_{05}$ & **** & 1810 - 1830 & $\frac{5}{2}^-$ & $1^-$ & $\frac{3}{2}$ & 1   & $\{n,s\}n$ & ${\bf 8}$ & ${\bf {6}}$      & 0 & $\frac{1}{2}$ & 0 & 1785  \\
$\Lambda(1890)$ $P_{03}$ & **** & 1850 - 1910 & $\frac{3}{2}^+$ & $0^+$ & $\frac{3}{2}$ & 1   & $\{n,s\}n$ & ${\bf 8}$ & ${\bf {6}}$      & 0 & $\frac{1}{2}$ & 0 & 1896  \\  
 missing                 & --   & --          & $\frac{1}{2}^+$ & $0^+$ & $\frac{1}{2}$ & 1   & $\{n,s\}n$ & ${\bf 8}$ & ${\bf {6}}$      & 0 & $\frac{1}{2}$ & 0 & 1955  \\ 
 missing                 & --   & --          & $\frac{1}{2}^+$ & $0^+$ & $\frac{1}{2}$ & 0   & $[n,s]n$   & ${\bf 8}$ & ${\bf {\bar 3}}$ & 0 & $\frac{1}{2}$ & 1 & 1960  \\
 missing                 & --   & --          & $\frac{1}{2}^-$ & $1^-$ & $\frac{1}{2}$ & 1   & $\{n,s\}n$ & ${\bf 8}$ & ${\bf {6}}$      & 0 & $\frac{1}{2}$ & 0 & 1969  \\
 missing                 & --   & --          & $\frac{3}{2}^-$ & $1^-$ & $\frac{1}{2}$ & 1   & $\{n,s\}n$ & ${\bf 8}$ & ${\bf {6}}$      & 0 & $\frac{1}{2}$ & 0 & 1969  \\  \\
$\Lambda^*(1405)$ $S_{01}$ & **** & 1402 - 1410 & $\frac{1}{2}^-$ & $1^-$ & $\frac{1}{2}$ & 0 & $[n,n]s$   & ${\bf 1}$ & ${\bf {\bar 3}}$ & 0 & 0             & 0 & 1431  \\
$\Lambda^*(1520)$ $D_{03}$ & **** & 1519 - 1521 & $\frac{3}{2}^-$ & $1^-$ & $\frac{1}{2}$ & 0 & $[n,n]s$   & ${\bf 1}$ & ${\bf {\bar 3}}$ & 0 & 0             & 0 & 1431  \\
 missing                 & --     & --          & $\frac{1}{2}^-$ & $1^-$ & $\frac{1}{2}$ & 0 & $[n,s]n$   & ${\bf 1}$ & ${\bf {\bar 3}}$ & 0 & $\frac{1}{2}$ & 0 & 1443  \\
 missing                 & --     & --          & $\frac{3}{2}^-$ & $1^-$ & $\frac{1}{2}$ & 0 & $[n,s]n$   & ${\bf 1}$ & ${\bf {\bar 3}}$ & 0 & $\frac{1}{2}$ & 0 & 1443  \\
  missing                 & --     & --         & $\frac{1}{2}^-$ & $1^-$ & $\frac{1}{2}$ & 0 & $[n,n]s$  & ${\bf 1}$ & ${\bf {\bar 3}}$ & 0 & 0               & 1 & 1854  \\
	  missing                 & --     & --         & $\frac{3}{2}^-$ & $1^-$ & $\frac{1}{2}$ & 0 & $[n,n]s$  & ${\bf 1}$ & ${\bf {\bar 3}}$ & 0 & 0               & 1 & 1854  \\
  missing                 & --     & --         & $\frac{1}{2}^-$ & $1^-$ & $\frac{1}{2}$ & 0 & $[n,s]n$  & ${\bf 1}$ & ${\bf {\bar 3}}$ & 0 & $\frac{1}{2}$   & 1 & 1928  \\ 
  missing                 & --     & --         & $\frac{3}{2}^-$ & $1^-$ & $\frac{1}{2}$ & 0 & $[n,s]n$  & ${\bf 1}$ & ${\bf {\bar 3}}$ & 0 & $\frac{1}{2}$   & 1 & 1928  \\  \\
\hline
\hline
\end{tabular}
\caption{As table \ref{tab:Spectrum-Sigma8}, but for $\Lambda$ and $\Lambda^*$-type resonances.} 
\label{tab:Spectrum-Lambda8}
\end{table*}
%%%%%%%%%%%%%%%%%%%%%%%%%%%%%%%%%%%%%%%%

\begin{appendix}

\section{Quark-diquark basis}
\label{qD basis}
For $N$-type states, we have the following states:
\begin{subequations}
\begin{equation}
	\left| [n,n]n; \mbox{ } \left(\mbox{{\boldmath{$\bar{3}$}}} \otimes \mbox{{\boldmath{$3$}}}\right) 
	\mbox{{\boldmath{$8$}}}; \mbox{ } \left(0,\frac{1}{2}\right)\frac{1}{2};
	\mbox{ } \left(0,\frac{1}{2}\right)\frac{1}{2} \right\rangle   \mbox{ },
\end{equation}
\begin{equation}
	\left| \{n,n\}n; \mbox{ } \left(\mbox{{\boldmath{$6$}}} \otimes \mbox{{\boldmath{$3$}}}\right) 
	\mbox{{\boldmath{$8$}}}; \mbox{ } \left(1,\frac{1}{2}\right)\frac{1}{2};
	\left(1,\frac{1}{2}\right)\frac{1}{2} \right\rangle   \mbox{ },
\end{equation}
\begin{equation}
	\left| \{n,n\}n; \mbox{ } \left(\mbox{{\boldmath{$6$}}} \otimes \mbox{{\boldmath{$3$}}}\right) 
	\mbox{{\boldmath{$8$}}}; \mbox{ } \left(1,\frac{1}{2}\right)\frac{1}{2};
	\left(1,\frac{1}{2}\right)\frac{3}{2} \right\rangle   \mbox{ };
\end{equation}
\end{subequations}
For $\Delta$-type states, one has:
\begin{subequations}
\begin{equation}
	\left| \{n,n\}n; \mbox{ } \left(\mbox{{\boldmath{$6$}}} \otimes \mbox{{\boldmath{$3$}}}\right) 
	\mbox{{\boldmath{$10$}}}; \mbox{ } \left(1,\frac{1}{2}\right)\frac{3}{2};
	\left(1,\frac{1}{2}\right)\frac{1}{2} \right\rangle   \mbox{ },
\end{equation}
\begin{equation}
	\left| \{n,n\}n; \mbox{ } \left(\mbox{{\boldmath{$6$}}} \otimes \mbox{{\boldmath{$3$}}}\right) 
	\mbox{{\boldmath{$10$}}}; \mbox{ } \left(1,\frac{1}{2}\right)\frac{3}{2};
	\left(1,\frac{1}{2}\right)\frac{3}{2} \right\rangle   \mbox{ };
\end{equation}
\end{subequations}
For $\Lambda$-type states one has:
\begin{subequations}
\begin{equation}
	\left| [n,n]s; \mbox{ } \left(\mbox{{\boldmath{$\bar{3}$}}} \otimes \mbox{{\boldmath{$3$}}}\right) 
	\mbox{{\boldmath{$8$}}}; \mbox{ } \left(0,0\right)0;
	\mbox{ } \left(0,\frac{1}{2}\right)\frac{1}{2} \right\rangle   \mbox{ },
\end{equation}
\begin{equation}
	\left| [n,s]n; \mbox{ } \left(\mbox{{\boldmath{$\bar{3}$}}} \otimes \mbox{{\boldmath{$3$}}}\right) 
	\mbox{{\boldmath{$8$}}}; \left(\frac{1}{2},\frac{1}{2}\right)0;
	\mbox{ } \left(0,\frac{1}{2}\right)\frac{1}{2} \right\rangle   \mbox{ },
\end{equation}
\begin{equation}
	\left| \{n,s\}n; \mbox{ } \left(\mbox{{\boldmath{$6$}}} \otimes \mbox{{\boldmath{$3$}}}\right) 
	\mbox{{\boldmath{$8$}}}; \mbox{ } \left(\frac{1}{2},\frac{1}{2}\right)0;
	\left(1,\frac{1}{2}\right)\frac{1}{2} \right\rangle   \mbox{ },
\end{equation}
\begin{equation}
	\left| \{n,s\}n; \mbox{ } \left(\mbox{{\boldmath{$6$}}} \otimes \mbox{{\boldmath{$3$}}}\right) 
	\mbox{{\boldmath{$8$}}}; \mbox{ } \left(\frac{1}{2},\frac{1}{2}\right)0;
	\left(1,\frac{1}{2}\right)\frac{3}{2} \right\rangle   \mbox{ };
\end{equation}
\end{subequations}
for $\Lambda^*$-type states one has:
\begin{subequations}
\begin{equation}
	\left| [n,n]s; \mbox{ } \left(\mbox{{\boldmath{$\bar{3}$}}} \otimes \mbox{{\boldmath{$3$}}}\right) 
	\mbox{{\boldmath{$1$}}}; \mbox{ } \left(0,0\right)0;
	\mbox{ } \left(0,\frac{1}{2}\right)\frac{1}{2} \right\rangle   \mbox{ },
\end{equation}
\begin{equation}
	\left| [n,s]n; \mbox{ } \left(\mbox{{\boldmath{$\bar{3}$}}} \otimes \mbox{{\boldmath{$3$}}}\right) 
	\mbox{{\boldmath{$1$}}}; \left(\frac{1}{2},\frac{1}{2}\right)0;
	\mbox{ } \left(0,\frac{1}{2}\right)\frac{1}{2} \right\rangle  \mbox{ };
\end{equation}
\end{subequations}
for $\Sigma$-type states one has:
\begin{subequations}
\begin{equation}
	\left| [n,s]n; \mbox{ } \left(\mbox{{\boldmath{$\bar{3}$}}} \otimes \mbox{{\boldmath{$3$}}}\right) 
	\mbox{{\boldmath{$8$}}}; \mbox{ } \left(\frac{1}{2},\frac{1}{2}\right)1;
	\mbox{ } \left(0,\frac{1}{2}\right)\frac{1}{2} \right\rangle   \mbox{ },
\end{equation}
\begin{equation}
	\left| \{n,n\}s; \mbox{ } \left(\mbox{{\boldmath{$6$}}} \otimes \mbox{{\boldmath{$3$}}}\right) 
	\mbox{{\boldmath{$8$}}}; \mbox{ } \left(1,0\right)1; \mbox{ } 
	\left(1,\frac{1}{2}\right)\frac{1}{2} \right\rangle   \mbox{ },
\end{equation}
\begin{equation}
	\left| \{n,n\}s; \mbox{ } \left(\mbox{{\boldmath{$6$}}} \otimes \mbox{{\boldmath{$3$}}}\right) 
	\mbox{{\boldmath{$8$}}}; \mbox{ } \mbox{ } \left(1,0\right)1; \mbox{ } 
	\left(1,\frac{1}{2}\right)\frac{3}{2} \right\rangle   \mbox{ },
\end{equation}
\begin{equation}
	\left| \{n,s\}n; \mbox{ } \left(\mbox{{\boldmath{$6$}}} \otimes \mbox{{\boldmath{$3$}}}\right) 
	\mbox{{\boldmath{$8$}}}; \mbox{ } \left(\frac{1}{2},\frac{1}{2}\right)1; \mbox{ } 
	\left(1,\frac{1}{2}\right)\frac{1}{2} \right\rangle   \mbox{ },
\end{equation}
\begin{equation}
	\left| \{n,s\}n; \mbox{ } \left(\mbox{{\boldmath{$6$}}} \otimes \mbox{{\boldmath{$3$}}}\right) 
	\mbox{{\boldmath{$8$}}}; \mbox{ } \left(\frac{1}{2},\frac{1}{2}\right)1; \mbox{ } 
	\left(1,\frac{1}{2}\right)\frac{3}{2} \right\rangle   \mbox{ };
\end{equation}
\end{subequations}
for $\Sigma^*$-type states one has:
\begin{subequations}
\begin{equation}
	\left| \{n,n\}s; \mbox{ } \left(\mbox{{\boldmath{$6$}}} \otimes \mbox{{\boldmath{$3$}}}\right) 
	\mbox{{\boldmath{$10$}}}; \mbox{ } \left(1,0\right)1; \mbox{ }
	\left(1,\frac{1}{2}\right)\frac{1}{2} \right\rangle   \mbox{ } \mbox{ },
\end{equation}
\begin{equation}
	\left| \{n,n\}s; \mbox{ } \left(\mbox{{\boldmath{$6$}}} \otimes \mbox{{\boldmath{$3$}}}\right) 
	\mbox{{\boldmath{$10$}}}; \mbox{ } \left(1,0\right)1; \mbox{ }
	\left(1,\frac{1}{2}\right)\frac{3}{2} \right\rangle   \mbox{ } \mbox{ },
\end{equation}
\begin{equation}
	\left| \{n,s\}n; \mbox{ } \left(\mbox{{\boldmath{$6$}}} \otimes \mbox{{\boldmath{$3$}}}\right) 
	\mbox{{\boldmath{$10$}}}; \mbox{ } \left(\frac{1}{2},\frac{1}{2}\right)1; \mbox{ }
	\left(1,\frac{1}{2}\right)\frac{1}{2} \right\rangle   \mbox{ } \mbox{ },
\end{equation}
\begin{equation}
	\left| \{n,s\}n; \mbox{ } \left(\mbox{{\boldmath{$6$}}} \otimes \mbox{{\boldmath{$3$}}}\right) 
	\mbox{{\boldmath{$10$}}}; \mbox{ } \left(\frac{1}{2},\frac{1}{2}\right)1; \mbox{ }
	\left(1,\frac{1}{2}\right)\frac{3}{2} \right\rangle   \mbox{ } \mbox{ };
\end{equation}
\end{subequations}
for $\Xi$-type states one has:
\begin{subequations}
\begin{equation}
	\left| [n,s]s; \mbox{ } \left(\mbox{{\boldmath{$\bar{3}$}}} \otimes \mbox{{\boldmath{$3$}}}\right) 
	\mbox{{\boldmath{$8$}}}; \mbox{ } \left(\frac{1}{2},0\right)\frac{1}{2};
	\mbox{ } \left(0,\frac{1}{2}\right)\frac{1}{2} \right\rangle  \mbox{ },
\end{equation}
\begin{equation}
	\left| \{n,s\}s; \mbox{ } \left(\mbox{{\boldmath{$6$}}} \otimes \mbox{{\boldmath{$3$}}}\right) 
	\mbox{{\boldmath{$8$}}}; \mbox{ } \left(\frac{1}{2},0\right)\frac{1}{2}; \mbox{ }
	\left(1,\frac{1}{2}\right)\frac{1}{2} \right\rangle  \mbox{ } \mbox{ },
\end{equation}
\begin{equation}
	\left| \{n,s\}s; \mbox{ } \left(\mbox{{\boldmath{$6$}}} \otimes \mbox{{\boldmath{$3$}}}\right) 
	\mbox{{\boldmath{$8$}}}; \mbox{ } \left(\frac{1}{2},0\right)\frac{1}{2}; \mbox{ }
	\left(1,\frac{1}{2}\right)\frac{3}{2} \right\rangle  \mbox{ } \mbox{ },
\end{equation}
\begin{equation}
	\left| \{s,s\}n; \mbox{ } \left(\mbox{{\boldmath{$6$}}} \otimes \mbox{{\boldmath{$3$}}}\right) 
	\mbox{{\boldmath{$8$}}}; \mbox{ } \left(0,\frac{1}{2}\right)\frac{1}{2}; \mbox{ }
	\left(1,\frac{1}{2}\right)\frac{1}{2} \right\rangle  \mbox{ } \mbox{ },
\end{equation}
\begin{equation}
	\left| \{s,s\}n; \mbox{ } \left(\mbox{{\boldmath{$6$}}} \otimes \mbox{{\boldmath{$3$}}}\right) 
	\mbox{{\boldmath{$8$}}}; \mbox{ } \left(0,\frac{1}{2}\right)\frac{1}{2}; \mbox{ }
	\left(1,\frac{1}{2}\right)\frac{3}{2} \right\rangle   \mbox{ } \mbox{ };
\end{equation}
\end{subequations}
for $\Xi^*$-type states one has:
\begin{subequations}
\begin{equation}
	\left| \{n,s\}s; \mbox{ } \left(\mbox{{\boldmath{$6$}}} \otimes \mbox{{\boldmath{$3$}}}\right) 
	\mbox{{\boldmath{$10$}}}; \mbox{ } \left(\frac{1}{2},0\right)\frac{1}{2}; \mbox{ }
	\left(1,\frac{1}{2}\right)\frac{1}{2} \right\rangle   \mbox{ } \mbox{ },
\end{equation}
\begin{equation}
	\left| \{n,s\}s; \mbox{ } \left(\mbox{{\boldmath{$6$}}} \otimes \mbox{{\boldmath{$3$}}}\right) 
	\mbox{{\boldmath{$10$}}}; \mbox{ } \left(\frac{1}{2},0\right)\frac{1}{2}; \mbox{ }
	\left(1,\frac{1}{2}\right)\frac{3}{2} \right\rangle   \mbox{ } \mbox{ },
\end{equation}
\begin{equation}
	\left| \{s,s\}n; \mbox{ } \left(\mbox{{\boldmath{$6$}}} \otimes \mbox{{\boldmath{$3$}}}\right) 
	\mbox{{\boldmath{$10$}}}; \mbox{ } \left(0,\frac{1}{2}\right)\frac{1}{2}; \mbox{ }
	\left(1,\frac{1}{2}\right)\frac{1}{2} \right\rangle   \mbox{ } \mbox{ },
\end{equation}
\begin{equation}
	\left| \{s,s\}n; \mbox{ } \left(\mbox{{\boldmath{$6$}}} \otimes \mbox{{\boldmath{$3$}}}\right) 
	\mbox{{\boldmath{$10$}}}; \mbox{ } \left(0,\frac{1}{2}\right)\frac{1}{2}; \mbox{ }
	\left(1,\frac{1}{2}\right)\frac{3}{2} \right\rangle   \mbox{ } \mbox{ };
\end{equation}
\end{subequations}
finally for $\Omega$-type states the only possibility is:
\begin{equation}
	\left| \{s,s\}s; \mbox{ } \left(\mbox{{\boldmath{$6$}}} \otimes \mbox{{\boldmath{$3$}}}\right) 
	\mbox{{\boldmath{$10$}}}; \mbox{ } \left(0,0\right)0; \mbox{ }
	\left(1,\frac{1}{2}\right)\frac{3}{2} \right\rangle   \mbox{ } \mbox{ }.
\end{equation}

\end{appendix}

%%%%%%%%%%%%%%%%%%%%%%%%%%%%%%%%%%%%%%%%%%%%%%%%%%

\end{document}